\documentstyle[aps,pra,epsfig,manuscript]{revtex}

\begin{document}
\title{Generation of dipole squeezing in a two-mode system with 
entangled coherent states of a quantized electromagnetic field} 
\author{Qin Rao} 
\address{Department of Engineering Physics, Queen's University, Kingston, 
ON K7L 3N6, Canada}
\author{Rui-Hua Xie} 
\address{Department of Chemistry, Queen's University, Kingston,
ON K7L 3N6, Canada}
\date{\today}
\maketitle
\begin{abstract}
Two-mode quantized electromagnetic fields can be entangled and
admit a large number of coherent states. In this paper, we 
consider a two-mode system that consists of a two-level atom 
interacting with  a two-mode quantized electromagnetic field, 
which is initially prepared in an entangled two-mode coherent 
state, via a nondegenerate two-photon  process in a lossless  
cavity. We study the quantum fluctuations in the two-mode system and 
investigate in detail the effects of detuning, Stark shift and 
atomic coherence on atomic dipole squeezing (ADS). We show that  ADS strongly 
depends on the atomic coherence.  It is found that the stronger the  
correlations between the two modes are involved,  the more the ADS 
could be generated. The detuning or Stark shift has a destructive 
effect on ADS, but the combined effect of the detuning and Stark shift 
may lead to a regular, periodical and strong  ADS pattern. 


\noindent
{\bf PACS (numbers)}: 42.50.Dv, 42.50.Lc, 03.67.-a

\end{abstract}

\newpage

\section{INTRODUCTION}

During the last few years, it has been shown that quantum entanglement
of states \cite{special0,paolo,tj00,af98,grover,wuxh96,chb92} 
is considered to be a basic prerequisite for teleporation of 
quantum state, quantum communication, 
quantum computing, quantum information processing and experimentally 
testing quantum mechanics. Theoretical and experimental studies on 
quantum communication through the use of quantum entangled states 
have shown great promise for establishing new kinds of 
information processing with no classical counterparts. One example is 
the concept of quantum teleporation\cite{af98}, where an entangled 
source was built from two single-mode phase squeezed vacuum states combined at 
a beamsplitter. Other examples of the application of entanglement
states to quantum communication include dense coding\cite{chb92} and 
cryptography\cite{tj00}. It is believed that the use of entangled states 
can lead to a substantial decrease of computational resources\cite{grover}. 
Also, entangled states have been widely used to test fundamental 
quantum features such as nonlocality\cite{wuxh96}. These extensive 
studies have paved a way toward a variety of useful quantum 
informational devices. 

Two-mode quantized electromagnetic fields can be entangled and 
admit a large number of coherent states\cite{ccg00}. Most of these 
coherent states may be associated with low-order Lie algebra su(1,1)
\cite{ccg00,puribook}. For example, two interesting types of su(1,1) 
coherent states are the Perelomov coherent states (PCS)
 \cite{ccg00,perelomov,ccg91} and Barut-Girardello coherent 
states\cite{ccg00,barut}. The former involves strong 
correlations between the modes and is generated from the unitary 
evolution of two-mode number states driven by a nondegenerated 
parametric device\cite{ccg00} (for example, the type that gives rise to 
down-conversion or harmonic generation). The latter type also 
involves tight correlations with respect to the photon number 
states between the modes and may be generated in a process involving the 
competition between nondegenerate parametric amplification and 
nondegenerate two-photon absorption\cite{ccg00}. The other class of two-mode 
su(1,1) coherent states is the intelligent states \cite{ccg951}, 
which equalize various possible uncertainty product that can be constructed 
as consequence of the su(1,1) commutation relation. Also, the two-mode 
squeezed pair-coherent state\cite{ccg952} is  an intelligent state but 
with unequal uncertainty. All of these two-mode su(1,1) coherent states 
may be written as superposition of the form 
$\sum_{n=0}^{\infty}C_{n}^{q}\mid n+q\rangle_{1}\mid n\rangle_{2}$ 
\cite{ccg00}, where the modes are labeled by 1 and 2, 
respectively, and the integer number 
$q$ is the (fixed) difference in the photon numbers of the two-modes and the 
coefficient $C_{n}^{q}$ depends on the specific su(1,1) coherent state. 
These states are of interest because they often display nonclassical 
characteristics such as squeezed light and photon antibunching as well as 
violations of the Cauchy-Schwarz inequality and Bell's inequality
\cite{ccg91,mdr86}. In this paper, we study further the properties 
of these correlated states of the field  since the importance lies in their
close connection to the two-photon nonlinear optical processes. 
On the other hand, because the study of the squeezed states can provide a 
fundamental understanding of quantum fluctuations and open the 
way for new schemes of quamtum communication or imaging beating 
the standard quantum noise limit,  we pay attention to investigating 
the generation of atomic dipole squeezing (ADS)\cite{dfw} in a two-mode 
system that consists of  a two-level atom interacting with a  
quantized electromagnetic field, which is initially prepared 
in an entangled  two-mode coherent state, via a nondegenerate two-photon 
process in a lossless cavity. We find that entangled two-mode coherent 
states of the field can also display ADS and that  the stronger the 
correlations between the modes are involved, the more the ADS could be 
generated.

The organization of this paper is as follows. 
In section II, we describe the theoretical 
model of the two-mode system and give its general solutions. 
In section III, we briefly introduce the definition of 
atomic dipole squeezing. In section IV, 
assuming that the two-level atom is  initially prepared in a coherent 
superposition of its ground and excited states and that 
the quantized electromagnetic field is initially in a 
two-mode PCS,  we investigate the generation of ADS in the 
two-mode system and examine in detail the effects of detuning, 
Stark shift and atomic coherence on ADS.  Finally, our conclusion and 
discussions are presented in section V. 

\section{THEORETICAL MODEL}

We consider a two-level atom interacting with two modes of 
a quantized electromagnetic field via a nondegenerate two-photon process 
in a lossless  cavity. For convenience, we denote 
$\omega$ as the natural transition frequency between the excited 
state $\mid+\rangle$ and the ground state $\mid-\rangle$ of the 
two-level atom and $\Omega_{j}$ as the frequency of mode $j$ of the 
field.  It is known that some intermediate virtual states 
$\mid {\bf virtual} \rangle$, which are assumed to be coupled to the 
excited  and ground states of the two-level atom by a dipole-allowed 
transition, are involved in the two-photon process. We assume that 
$\Omega_{1}$ is about several times $\Omega_{2}$ and the two modes 
are tuned as near one-photon resonance as possible but 
still off the one-photon linewidth, i.e., the detuning 
$\delta_{1}$ and $\delta_{2}$ are larger than but as close as 
the linewidth of $\mid {\bf virtual} \rangle$. For the case of 
$\delta_{1}=-\delta_{2}=\delta$ and by the method of adiabatic 
elimination of the virtual state $\mid {\bf virtual} \rangle$
\cite{alsing}, Gou\cite{gou90} obtained an effective 
Hamiltonian for this system within a rotating-wave approximation 
\begin{equation}
H=\Omega_{1}a_{1}^{\dagger}a_{1} 
+\Omega_{2}a_{2}^{\dagger}a_{2}+\omega S_{z}+
\beta_{2}S_{+}S_{-}a_{2}^{+}a_{2}+\beta_{1}S_{-}S_{+}a_{1}^{\dagger}a_{1}
+\epsilon (a_{1}^{\dagger}a_{2}^{\dagger}S_{-}+a_{1}a_{2}S_{+}),
\end{equation}
where $\epsilon=\sqrt{\beta_{1}\beta_{2}}$. Here  
$S_{z}$ and $S_{\pm}$ are operators of
the atomic pseudospin inversion and transition, respectively, 
and satisfy the commutation
\begin{equation}
[S_{+}, S_{-} ] = 2 S_{z},
\end{equation}
\begin{equation}
[S_{z}, S_{\pm} ] = \pm S_{\pm}.
\end{equation}
$a_{i}^{\dagger}$ and $a_{i}$ are the creation and 
annihilation operators for the
photons of mode $i$, 
respectively,  which obey the boson operators' 
commutation relation
\begin{equation}
[a_{i}, a_{j}^{\dagger}]=\delta_{ij}.
\end{equation}
$\epsilon$ is the
coupling constant between the atom and the  field.
A Stark shift is caused by the intermediate virtual state 
 $\mid {\bf virtual} \rangle$, 
 and the corresponding parameter $\beta_{1}$ and $\beta_{2}$ are  given 
by $\beta_{1}=\epsilon_{1}^{2}/\delta$, 
$\beta_{2}=\epsilon_{2}^{2}/\delta$ 
and $\epsilon=\epsilon_{1}\epsilon_{2}/\delta$, where 
$\delta=\omega-(\Omega_{1}+\Omega_{2})$ and 
$\epsilon_{1}$ and $\epsilon_{2}$ are the coupling strengths of 
the intermediate virtual state with the ground and excited states of the 
two-level atom, respectively. Throughout we employ the units of 
$\hbar=c=1$.

We denote  $\mid n\rangle$ as the Fock state of the field. Using 
the standard bare-state procedure introduced by
Yoo and Eberly\cite{yoo}, we are able
to calculate  the expectation value
of any operator in the two-mode system described above. 
In an ideal cavity, the two-level 
atom absorbs and emits one pair of photons, and the two bases in the
bare-state representation are given by 
$\mid +, n_{1}, n_{2}\rangle$ and $\mid -, n_{1}+1, n_{2}+1\rangle$, 
respectively. Thus, the total Hamiltonian given by Eq.(1) can be written as:
\begin{equation}
H=\left[\begin{array}{ccc} \Omega_{1}n_{1}
+\Omega_{2}n_{2}+\omega/2+\beta_{2}n_{2} &
\ \ \  &
\epsilon\sqrt{(n_{1}+1)(n_{2}+1)}\\
\epsilon\sqrt{(n_{1}+1)(n_{2}+1)} & \ \ \  &
\Omega_{1}(n_{1}+1)+\Omega_{2}(n_{2}+1)-\omega/2+\beta_{1}(n_{1}+1)
\end{array}\right].
\end{equation}
The eigenvalues of the above Hamiltonian are 
\begin{eqnarray}
\sigma_{+}&=&\Omega_{1}n_{1}+\Omega_{2}n_{2}+\frac{1}{2}[\omega+ 
\beta_{2}n_{2}+\beta_{1}(n_{1}+1)]+\epsilon\Gamma(n_{1},n_{2}), \\
\sigma_{-}&=&\Omega_{1}n_{1}+\Omega_{2}n_{2}+\frac{1}{2}[\omega+
\beta_{2}n_{2}+\beta_{1}(n_{1}+1)]-\epsilon\Gamma(n_{1},n_{2})
\end{eqnarray}
with 
\begin{eqnarray}
\Gamma(n_{1},n_{2})&=&\sqrt{\chi^{2}(
\mu,n_{1},n_{2})/4+(n_{1}+1)(n_{2}+1)},\\
\chi(\mu,n_{1},n_{2})&=&\left\{\begin{array}{ccc} 
\Delta+[\mu (n_{1}+1)-n_{2}]/\sqrt{\mu} &\  &\mu\ne 0, \\
\Delta & \  & \mu =0, \end{array}\right.
\end{eqnarray}
where $\mu=\beta_{1}/\beta_{2}$ and 
 $\Delta=\delta/\epsilon$ is the scaled detuning  
and the two-photon Rabi frequency is  written as 
\begin{equation}
\Omega(n_{1},n_{2})=\epsilon\Gamma(n_{1},n_{2}).
\end{equation}
It should be mentioned that $\mu=0$ is the case in the absence of
Stark shift. Then, the time evolution operator $U(n_{1},n_{2},t)$
of the two-mode system is given by  
\begin{equation}
U(n_{1},n_{2},t)=
\frac{1}{2}\left[\begin{array}{cc} E_{+}(t) & \zeta(t) \\
\zeta(t) & E_{-} (t)
  \end{array}\right],
\end{equation}
where
\begin{eqnarray}
E_{+}(t)&=&\left[1+
\frac{\chi(\mu,n_{1},n_{2})}{2\Gamma(n_{1},n_{2})}
\right]\exp(-i\sigma_{+}t)+
\left[1-\frac{\chi(\mu,n_{1},n_{2})}{2\Gamma(n_{1},n_{2})}
\right]\exp(-i\sigma_{-}t),\nonumber\\
E_{-}(t)&=&\left[1-\frac{\chi(\mu,n_{1},n_{2})}{2\Gamma(n_{1},n_{2})}
\right]\exp(-i\sigma_{+}t)+\left[1
+\frac{\chi(\mu,n_{1},n_{2})}{2\Gamma(n_{1},n_{2})}
\right]\exp(-i\sigma_{-}t), \nonumber\\
\zeta(t)&=&\sqrt{(n_{1}+1)(n_{2}+1)}
\left[\exp(-i\sigma_{+}t)-\exp(-i\sigma_{-}t)\right]/\Gamma(n_{1},n_{2}).
\end{eqnarray}
In the resonant case (i.e, $\Delta=0$), our solutions are in agreement 
with those obtained by Gou\cite{gou90,gou89}. Once the matrix 
representation of $U(n_{1}, n_{2}, t)$ is obtained, the density 
operator of the two-mode  system at time $t$ 
can be calculated with an arbitrary initial condition $\rho(t=0)$ by
\begin{equation}
\rho(t)=U(n_{1}, n_{2}, t)\rho(t=0)U^{\dagger}(n_{1}, n_{2}, t).
\end{equation}
Thus, the expectation value of any physical operator $O$ and its 
dependence on time  can be obtained through the formula 
\begin{equation}
\langle O(t)\rangle = Tr[\rho(t)O(0)].
\end{equation}

\section{DEFINITION OF ATOMIC DIPOLE SQUEEZING}

Squeezed states\cite{special0,xiepra02,app1} are distinguished by
the property that the quantum fluctuations in a dynamical
observable may be reduced below the standard quantum
noise limit at the expense of increased fluctuations in its
canonical conjugated variable without violating the
Heisenberg uncertain relation. Since the first observation 
of squeezed states of light, the generation of squeezed states 
has been the subject of an intense theoretical and 
experimental activity \cite{special0,book00,book01,special92}. 
The study of these states provides a fundamental understanding 
of quantum fluctuations and opens a way for new schemes 
\cite{special0,book00,book01,special92} of quantum communication or 
imaging beating the standard quantum noise limit.  It has been  
predicted that a number of physical systems
\cite{special0,book00,book01,special92,xie02,xie022} 
(e.g.,  four-wave mixing, two-photon laser, parametric 
amplifiers and Raman scattering)  can generate squeezed states of 
 the radiation field. Also, it is known that atomic dipole squeezing 
(ADS)\cite{dfw}  could be generated through many kinds of 
nonlinear optical processes\cite{dfw,xsli89,ref6,ref7,xie95}, such as
 resonance fluorescence, cooperative Dicke system,
Rydberg atom maser, and so on. For the special case of a
vacuum-field mode, Wodkiewicz et al.\cite{ref9} examined
the relationship between ADS and field squeezing. Very recently, 
a general relationship between field squeezing 
and ADS has been established under different initial conditions
for both the field and atom in a high-Q micromaser cavity 
\cite{ref133,ref13,ref12,ref14,ref11,ref122}, and it has been 
shown that squeezed atom can generate squeezed light. 
In this paper, assuming that the field is initially prepared 
in an entangled  two-mode coherent states, we study the 
generation of ADS in the two-mode system introduced above.   

Here we briefly introduce the definition of atomic dipole squeezing
\cite{dfw}. Following the standard procedure, we define two slowly 
varying Hermitian quadrature operators\cite{xie95}
\begin{eqnarray}
S_{x}&=&\frac{1}{2}\left[ S_{+}\exp(-i\omega t)+S_{-}
\exp(i\omega t)\right],\\
S_{y}&=&\frac{1}{2i}\left[S_{+}\exp(-i\omega t)-S_{-}
\exp(i\omega t)\right],
\end{eqnarray}
where $S_{x}$ and $S_{y}$, actually, correspond to the dispersive and
absorptive components of the slowly varying atomic dipole\cite{dipole}, 
respectively. They  obey a commutation relation given by 
$[S_{x}, S_{y}] = i S_{z}$. Correspondingly, the Heisenberg 
uncertainty relation is\cite{xie95}
\begin{equation}
(\Delta S_{x})^{2}(\Delta S_{y})^{2} \ge \frac{1}{4}
\left|\langle S_{z}\rangle\right|^{2},
\end{equation}
where $(\Delta S_{j})^{2}=\langle S_{j}^{2}\rangle -
\langle S_{j}\rangle^{2}$  is the variance in the quadrature component 
$S_{j}$ (j=x,y) of the atomic dipole.  For convenience, we define 
the following functions\cite{xie95}
\begin{eqnarray}
F_{1}&=&(\Delta S_{x})^{2}-\frac{1}{2}\left|\langle S_{z}\rangle\right|, \\
F_{2}&=&(\Delta S_{y})^{2}-\frac{1}{2}\left|\langle S_{z}\rangle\right|.
\end{eqnarray}
Then, the atomic state is said to be squeezed if $F_{1}<0$ (or $F_{2}<0$).
This is the general definition of ADS\cite{dfw,xsli89,ref6,ref7,xie95}.

\section{Results}

From the two-photon Rabi frequency $\Omega(n_{1},n_{2})$ given by
Eq.(10), we find that one of its terms, 
 $\chi(\mu,n_{1},n_{2})$ given by Eq.(9), 
would not vanish for general two-mode field states
due to the arbitrariness of photon number $n_{1}$
and $n_{2}$ which are presented in the term $\chi(\mu,n_{1},n_{2})$. 
Hence, in this sense, we cannot ignore the effect of the Stark shift 
in the two-mode system. Of course, it is possible to 
eliminate or weaken this effect. For example, if we take $\mu=1$, 
then we have 
\begin{equation}
\chi(\mu=1,n_{1},n_{2})=\Delta+n_{1}-n_{2}+1.
\end{equation}
 It is obvious that if there exists some
kinds of correlation between the two-mode fields, for example,
the difference between their photon numbers is fixed
at a constant, i.e., $n_{1}-n_{2}=q$ ($q$, an invariant integer), then
the term $\chi$ would be fixed at $\chi=\Delta+q+1$ which is
intensity independent. Moreover, if we choose the case of
$q=-1-\Delta$ (assuming that the scaled detuning $\Delta$ takes an 
integer value), we see that the term $\chi$ would vanish. 
This is only true for physical quantities which are not sensitive 
to the phase\cite{gou90}. For phase-sensitive quantities, the 
Stark shift still plays an important role in determining their 
dynamical behaviours. We shall clarify this point in our following 
results. 

In the two-mode system,  we assume that the two-level atom is 
initially prepared in a coherent superposition of its 
ground state $\mid -\rangle$ and excited 
state $\mid +\rangle$\cite{xie95} 
\begin{equation}
\mid atom\rangle=c_{-}\mid -\rangle+c_{+}\mid+\rangle 
\end{equation}
 and that the two-mode field is initially in a two-mode PCS 
given by\cite{ccg00} 
\begin{equation}
\mid field\rangle\equiv\sum_{n=0}^{\infty}
(1-\mid\xi\mid^{2})^{(1+q)/2}\left[
\frac{(n+q)!}{n!q!}\right]^{1/2}\xi^{n} 
\mid n+q\rangle_{1}\mid n\rangle_{2}, 
\end{equation}
where $\xi=-\tanh(\theta/2)\exp(-i\phi)\equiv \mid\xi\mid\exp(i\Phi_{2})$ with
$0< \mid\xi\mid < 1$ and the phase $\Phi_{2}$, and the number $q$ 
(a positive integer or zero) is the
eigenvalue of $\mid a_{1}^{\dagger}a_{1}-a_{2}^{\dagger}a_{2}\mid$. 
Using Eqs.(14), we arrive at the expectation values of atomic inversion 
$S_{z}$, the quadrature components $S_{x}$ and $S_{y}$ of atomic 
dipole 
\begin{eqnarray}
\langle S_{z}(t)\rangle &=&\frac{1}{2}\sum_{n=0}^{\infty}
P(n,q)\left\{ 
\mid c_{+}\mid^{2}\left[
\cos^{2}[\nu(n,q)\tau]+\left(\frac{V^{2}(n,q)}{4}
-U^{2}(n,q)\right)\sin^{2}[\nu(n,q)\tau]\right]\right.\nonumber\\
 &-&\mid c_{-}\mid^{2}\left[
\cos^{2}[\nu(n-1,q)\tau]+\left(\frac{V^{2}(n-1,q)}{4}
-U^{2}(n-1,q)\right)\sin^{2}[\nu(n-1,q)\tau]\right]\nonumber\\
&-&4\mid c_{+}\mid\mid c_{-}\mid \Pi(n,q)U(n,q)\times 
\nonumber\\ 
& &\left.\left(\cos[\nu(n,q)\tau]\sin(\Phi_{1}-\Phi_{2})-
\frac{V(n,q)}{2}\sin[\nu(n,q)\tau]\cos(\Phi_{1}-\Phi_{2})
\right)\sin[\nu(n,q)\tau]\right\},
\end{eqnarray}

\begin{eqnarray}
\langle S_{x}(\tau)\rangle&=&\sum_{n=0}^{\infty}P(n,q)
\left\{\mid c_{+}\mid^{2}\Pi(n,q)U(n,q)\sin[\nu(n,q)\tau]\times 
\right.\nonumber\\ 
& &\left(\frac{V(n+1,q)}{2}
\sin[\nu(n+1,q)\tau]\cos(\Phi_{\tau}-\Phi_{2})+
\cos[\nu(n+1,q)\tau]\sin(\Phi_{\tau}-\Phi_{2})\right)\nonumber\\
&-&\mid c_{-}\mid^{2}\Pi(n,q)U(n,q)\sin[\nu(n,q)\tau]
\times\nonumber\\ 
& &\left(\frac{V(n-1,q)}{2}\sin[\nu(n-1,q)\tau]\cos(\Phi_{\tau}-\Phi_{2})
+\cos[\nu (n-1,q)\tau]\sin(\Phi_{\tau}-\Phi_{2})\right)\nonumber\\
&+&\mid c_{+}\mid\mid c_{-}\mid\left[ 
U(n,q)U(n+1,q)\Pi(n,q)\Pi(n+1,q)\times\right.\nonumber\\
& &\sin[\nu(n,q)\tau]\sin[\nu(n+1,q)\tau]\cos(\Phi_{\tau}+\Phi_{1}-2\Phi_{2})
+\cos(\Phi_{\tau}-\Phi_{1})\times\nonumber\\ 
& &\left(\cos[\nu(n-1,q)\tau]\cos[\nu(n,q)\tau]-\frac{V(n,q)V(n-1,q)}{4}
\sin[\nu(n-1,q)\tau]\sin[\nu(n,q)\tau]\right)\nonumber\\ 
&-&\sin(\Phi_{\tau}-\Phi_{1})\left(\frac{V(n,q)}{2}\cos[\nu(n-1,q)\tau]
\sin[\nu(n,q)\tau]\right.\nonumber\\ 
&+&\left.\left.\left.\frac{V(n-1,q)}{2}
\cos[\nu(n,q)\tau]\sin[\nu(n-1,q)\tau]\right)\right]\right\}, 
\end{eqnarray}

\begin{eqnarray}
\langle S_{y}(\tau)\rangle&=&\sum_{n=0}^{\infty}P(n,q)
\left\{\mid c_{+}\mid^{2}\Pi(n,q)U(n,q)\sin[\nu(n,q)\tau]\times
\right.\nonumber\\
& &\left(\frac{V(n+1,q)}{2}
\sin[\nu(n+1,q)\tau]\sin(\Phi_{\tau}-\Phi_{2})-
\cos[\nu(n+1,q)\tau]\cos(\Phi_{\tau}-\Phi_{2})\right)\nonumber\\
&-&\mid c_{-}\mid^{2}\Pi(n,q)U(n,q)\sin[\nu(n,q)\tau]
\times\nonumber\\
& &\left(\frac{V(n-1,q)}{2}\sin[\nu(n-1,q)\tau]\sin(\Phi_{\tau}-\Phi_{2})
-\cos[\nu (n-1,q)\tau]\cos(\Phi_{\tau}-\Phi_{2})\right)\nonumber\\
&+&\mid c_{+}\mid\mid c_{-}\mid\left[
U(n,q)U(n+1,q)\Pi(n,q)\Pi(n+1,q)\times\right.\nonumber\\
& &\sin[\nu(n,q)\tau]\sin[\nu(n+1,q)\tau]\sin(\Phi_{\tau}+\Phi_{1}-2\Phi_{2})
+\sin(\Phi_{\tau}-\Phi_{1})\times\nonumber\\
& &\left(\cos[\nu(n-1,q)\tau]\cos[\nu(n,q)\tau]-\frac{V(n,q)V(n-1,q)}{4}
\sin[\nu(n-1,q)\tau]\sin[\nu(n,q)\tau]\right)\nonumber\\
&+&\cos(\Phi_{\tau}-\Phi_{1})\left(\frac{V(n,q)}{2}\cos[\nu(n-1,q)\tau]
\sin[\nu(n,q)\tau]\right.\nonumber\\
&+&\left.\left.\left.\frac{V(n-1,q)}{2}
\cos[\nu(n,q)\tau]\sin[\nu(n-1,q)\tau]\right)\right]\right\}
\end{eqnarray}
with 
\begin{eqnarray}
U(n,q) &=&\sqrt{(n+1)(n+q+1)}/\nu(n,q),\\
V(n,q) &=&\chi(\mu,n_{1}=n+q,n_{2}=n)/\nu (n,q),\\
\nu(n,q)&=&\Gamma(n_{1}=n+q,n_{2}=n),\\
\Pi(n,q)&=&\mid\xi\mid \sqrt{(n+q+1)/(n+1)}\\ 
\Phi_{\tau} &=& \left\{
 \begin{array}{ccc} (\mu+1)\tau/(2\sqrt{\mu}) & \ & \mu\ne 0, \\
 0 & \ & \mu=0,\end{array}\right. \\
P(n,q) &=&[1-\mid\xi\mid^{2}]^{q+1}\mid\xi\mid^{2n}
(n+q)!/[n!q!] 
\end{eqnarray}
where $\tau=\epsilon t$ is the scaled interaction time, 
$\Phi_{1}$ is the initial relative phase between the ground  and 
excited states of the two-level atom,  and $\Phi_{\tau}$ is the 
induced phase due to the effect of Stark shift in the two-mode 
system.

Obviously, if we choose $\mu=1$ and $q=-1-\Delta$, then the term 
$\chi$ in the atomic inversion $S_{z}$ given by Eq.(23) 
vanishs. Thus the effect of Stark shift on $S_{z}$ can be 
totally eliminated. However, for phase-sensitive quantities, 
 for example, the two components $S_{x}$ and $S_{y}$ of the atomic dipole 
given by Eq.(24) and (25), respectively, we see that 
the Stark-shift-induced phase $\Phi_{\tau}$ could not 
vanish with the time evolution of the two-mode system, i.e., 
the effect of Stark shift on $S_{x}$ or $S_{y}$ cannot be 
eliminated.   As an example,  Fig.1(a-c) show the time evolution 
of $S_{z}$ and $S_{y}$ for the case of $q=1$ when the two-level atom 
is initially prepared in its excited state. Comparing Fig.1(b) with  
Fig.1(a), one can see the distinguished effect of the Stark shift 
on the dynamical behaviours of $S_{z}$ and $S_{y}$. Moreover, 
comparing Fig.1(c) with Fig.1(a), one can find  that 
for the case of  $\Delta=-2$ and $\mu=1$ the effect of 
Stark shift on $S_{z}$ is totally eliminated, 
but that on  $S_{y}$ still exists. Hence, we may conclude that 
the presence of Stark shift could bring phase information into 
the quadrature components $S_{x}$ and $S_{y}$, but not to the atomic 
inversion $S_{z}$. Thus, we would  expect that there is a 
significant effect of Stark shift on the squeezing properties of 
quantum fluctuations in one of the quadrature components of the atomic
dipole. 

In the following, taking the quantum fluctuations in the 
quadrature component $S_{x}$ of the atomic dipole as an example, 
we investigate the generation of ADS in the two-mode system. Throughout, 
we set the phase of the parameter $\xi$ at zero, i.e., $\Phi_{2}=0$. 
First, we examine the resonant case and consider that the Stark shift 
is absent. The numerical results are shown in Fig.2-4. 

Assuming that the atom is initially prepared in a coherent superposition 
of its ground and excited states, as shown in Fig.2(a-c),  we see that 
regular and periodical ADS patterns are observed for the cases of 
$\mid\xi\mid=0.1$ and $q=1, 4, 8$. As the value of the parameter 
$\mid\xi\mid$ is increased to 0.5, we find that the regular and 
periodical ADS pattern is destroyed greatly. Moreover, comparing 
Fig.2(b)(c) with Fig.2(a), we find that as the number $q$ varies from 
1 to 8,  the squeezing duration is decreased gradually and the squeezing 
period is also shortened. In Fig.2(d), we display the relation 
between the function $F_{1}$ and the number $q$ at a fixed interaction 
time $\tau=0.8\pi$ for the case of $\mid\xi\mid=0.1$. It is
 found that the squeeze depth (strength) is decreased as the number $q$ 
 varies from 1 to 200 and ADS can be shown only for $q<30$. This implies that 
the stronger the correlations between the modes are involved, the more the 
atomic dipole squeezing could be generated.

Next, we examine the effect of atomic coherence on ADS exhibited in 
the two-mode system. Holding the phase $\Phi_{1}$ at zero 
and the interaction time $\tau$ at $0.8\pi$, we show in Fig.3 
the relation between $F_{1}$ and 
the coefficient $\mid c_{-}\mid^{2}$ for $\xi=0.1$ as the number $q$ varies 
from 1 to 8.  For $q=1$, it is seen that ADS can be generated 
in a wide squeezing range 
of the parameter $\mid c_{-}\mid^{2}$ (i.e., 
$0.5< \mid c_{-}\mid^{2} <1$). However, increasing the number 
$q$ would narrow down the squeezing range, and ADS depth 
is decreased greatly. 

Then, we turn to examine the phase dependence of ADS generated in the 
two-mode system. For $\xi=0.1$,  $\tau=0.8\pi$  and $\mid c_{-}\mid^{2}=0.8$, 
 Fig.4 presents the relation between $F_{1}$ and the relative phase 
$\Phi_{1}$  as the number $q$ varies from 1 to 8. We find that in the case 
of $q=1$ ADS can be exhibited in a wide squeezing range of $\Phi_{1}$ around 
$0, \pi, 2\pi$. However, the squeezing range is narrowed down as the number 
$q$ increases. 

Finally, we consider the nonresonant case and examine in detail 
the effects of detuning and/or  Stark shift on ADS. The numerical 
results are shown in Fig.5(a-d) for q=1, $\mid\xi\mid=0.1$ and 
different choices of parameter pair $\Delta$ and $\mu$. 
Comparing Fig.5(b)(d)  with Fig. 5(a), we find that the detuning or 
Stark shift has a destructive effect on the ADS pattern exhibited 
in Fig.5(a). The squeezing duration is decreased and the ADS pattern 
is irregular and nonperiodic. For the case of $\Delta=-2$ and $\mu=1$, 
as discussed  above, the effects of Stark shift on $S_{z}$ can 
be eliminated totally but those on $S_{x}$ and $S_{y}$ 
cannot. Hence, we would expect a significant effect of the Stark shift 
on ADS. As shown in Fig.5(c), we really observe a regular, periodic and 
strong  ADS  pattern which is generated due to the combined effect of 
detuning and Stark shift. Compared to Fig.5(a), the squeezing duration  
in Fig.5(c) is decreased due to the combined effect of detuning and 
Stark shift.

\section{CONCLUSION AND DISCUSSION}

In this paper, we have considered a two-mode system that consists of a 
two-level atom interacting with  a quantized electromagnetic 
field, which is initially
prepared in an entangled two-mode PCS, via a nondegenerate two-photon 
process in a lossless  cavity. We have investigated the generation of 
ADS in this system and examine  in detail the effects of detuning, 
Stark shift and atomic coherence on ADS. We show that  ADS strongly 
depends on the atomic coherence.  It is found that the stronger the 
correlations between the modes are involved, the more the ADS can be 
generated. The detuning or Stark shift has a destructive effect on ADS. 
However, the combined effect of the detuning and Stark shift may lead 
to periodical, regular and strong ADS pattern but with decreased squeeze 
duration. 

Gou\cite{gou89} have shown that the correlation is responsible for 
generating the squeezing of the cavity fields. Our prediction of ADS 
in the two-mode system agrees well with this argument. In addition, 
it is obvious that two-mode squeezed vacuum\cite{ccg00} (i.e., the 
special case of the two-mode PCS for $q=0$) is a strongly correlated 
quantum state since in the number-state expansion there are 
always an equal number of photons in the $a_{1}$ and $a_{2}$ modes. 
Hence, a strong ADS pattern in the two-mode system would be expected 
if the field is initially prepared in a two-mode squeezed vacuum. This 
point can be clearly seen from Fig.6, where we compare the ADS 
pattern of $q=0$ with that of $q=1$. 
 
Actually, in this paper, our discussion is  restricted to an ideal 
case, i.e., without considering the cavity damping in the two-mode system. 
In the realistic situation, a high but finite cavity Q indeed causes 
some inevitable effects of photon dissipation in the cavity. Hence, it 
would be significant and interesting to include the cavity damping in the 
two-mode system and examine the effect of cavity  damping on ADS predicted 
in this paper.  

Finally, it should be mentioned that there are several significant 
experimental schemes which involve the nonresonant interaction of the 
atom with two nondegenerate cavity modes, for example, those 
illustrated in the works of Gou\cite{gou89} and 
Abdalla {\sl et al.} \cite{msa91}, and the recent one realized 
by Haroche's group\cite{pb02} with a single circular Rydberg atom in 
a cavity.  Certainly, these schemes could be helpful for realizing the 
two-mode system introduced in section II and for further understanding 
the ADS predicted in this paper.

\section*{ACKNOWLEDGEMENT} 

One of us (Q.R) thanks the Queen's  University for a Reinhart Fellowship.

\newpage


\begin{center} {\Large\bf CAPTION OF FIGURES }\end{center}

{\bf FIG.1}: Time evolution of $S_{z}$ and $S_{y}$ for $q=1$, 
$\mid\xi\mid=0.9$, $\Phi_{2}=0$, $\mid c_{-}\mid^{2}=0$ and  
(a) $\Delta=0$, $\mu=0$; (b) $\Delta=0$, $\mu=1$;
(c) $\Delta=-2$, $\mu=1$. 

\

{\bf FIG.2}: Time evolution of $F_{1}$ for $\Delta=0$, 
$\mu=0$, $\mid c_{-}\mid^{2}=0.8$ and 
$\Phi_{1}=0$  when the two-mode field is  in a two-mode PCS 
with (a) q=1, (b) q=4 and (c) q=8, where the solid and dot-dashed 
lines are for $\mid\xi\mid=0.1, 0.5$, respectively.  Case (d) is the
relation between $F_{1}$ and $q$ for $\mid\xi\mid=0.1$ at a 
fixed time $\tau=0.8\pi$. 

\

{\bf FIG.3}: Relation between $F_{1}$ and 
coefficient $\mid c_{-}\mid^{2}$ for 
$\Delta=0$, $\mu=0$, $\mid\xi\mid=0.1$ and $\Phi_{1}=0$ at 
a fixed time $\tau=0.8\pi$, where the solid, dashed and 
dot-dashed  lines are for $q=1, 4, 8$, respectively.

\
 
{\bf FIG.4}: Relation between $F_{1}$ and 
phase  $\Phi_{1}$ for $\Delta=0$ and $\mu=0$, 
$\mid\xi\mid=0.1$ and $\mid c_{-}\mid^{2}=0.8$ 
 at a fixed time $\tau=0.8\pi$, where the solid, 
dashed and dot-dashed lines are for $q=1, 4, 8$, 
respectively.

\

{\bf FIG.5}: Time evolution of $F_{1}$ for $\mid\xi\mid=0.1$, 
$\mid c_{-}\mid^{2}=0.8$ and $\Phi_{1}=0$ when the two-mode 
field is initially prepared in a two-mode PCS with q=1: 
(a) $\Delta=0$,  $\mu=0$;
(b) $\Delta=0$,  $\mu=1$;
(c) $\Delta=-2$, $\mu=1$;
(d) $\Delta=-2$, $\mu=0$.

\
 
{\bf FIG.6}: Time evolution of $F_{1}$ for $\Delta=0$, 
$\mu=0$, $\mid c_{-}\mid^{2}=0.8$, $\Phi_{1}=0$  and 
$\mid\xi\mid=0.1$ when  the two-mode field is  in a two-mode PCS, 
 where the dot-dashed and solid lines are for $q=0, 1$, 
 respectively.

\newpage 

\begin{center}{\Large\bf Rao \& Xie: FIG.1/PHYSA}\end{center}
 
\vspace{2.5cm}
 
\begin{center}
 
\epsfig{file=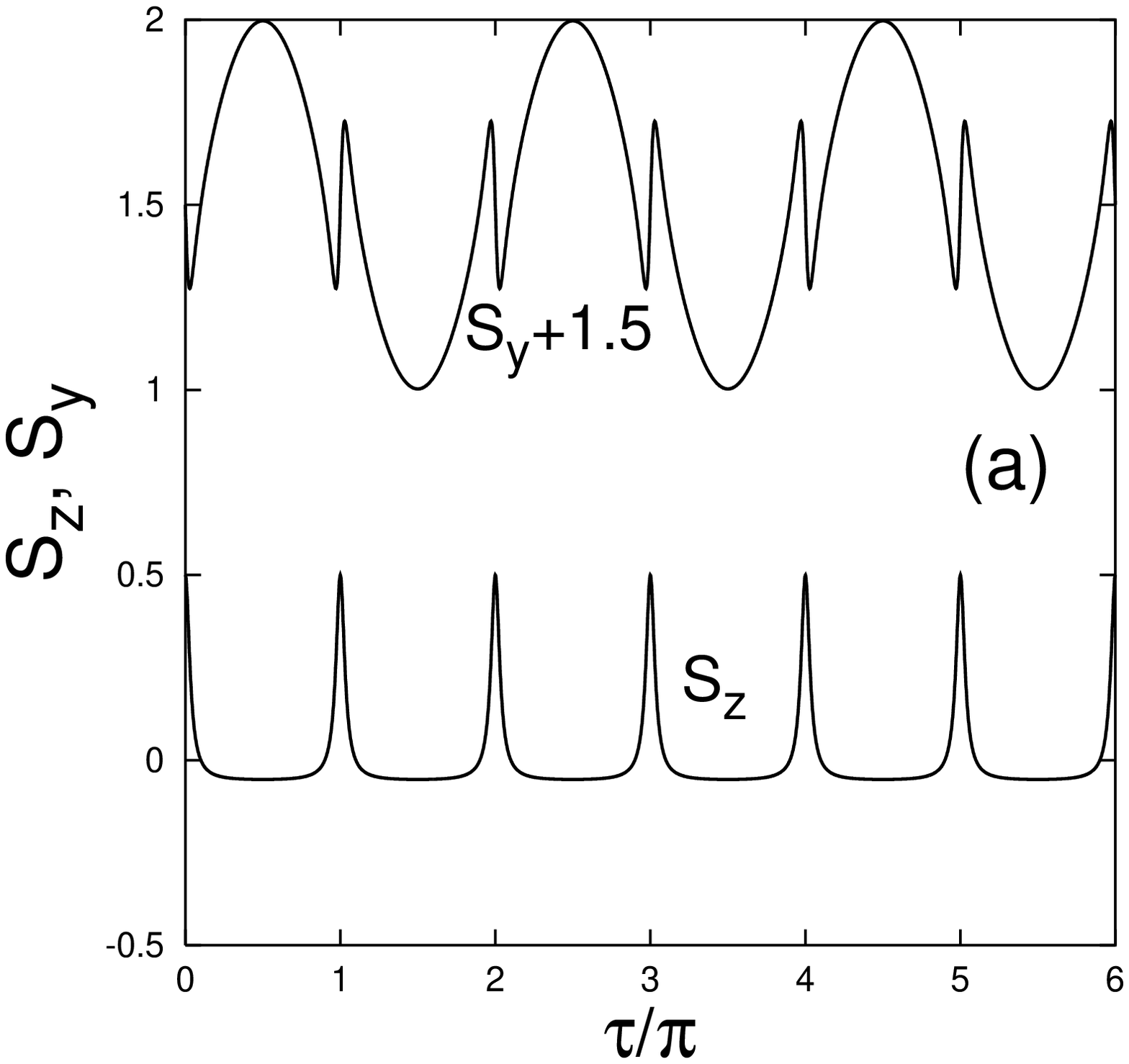,width=7.5cm,height=7.5cm}
\epsfig{file=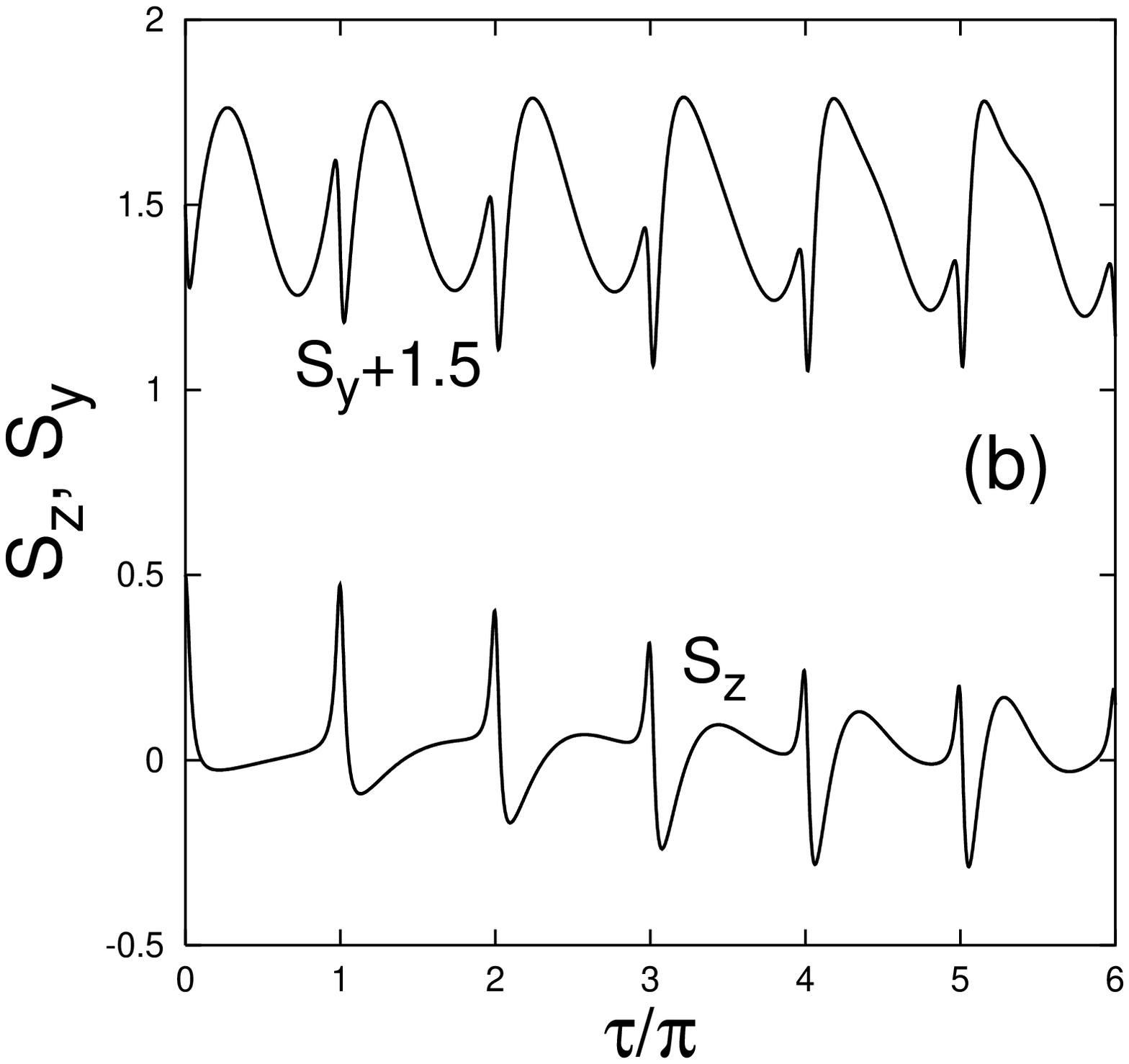,width=7.5cm,height=7.5cm}
 
\vspace{0.5cm}
 
\epsfig{file=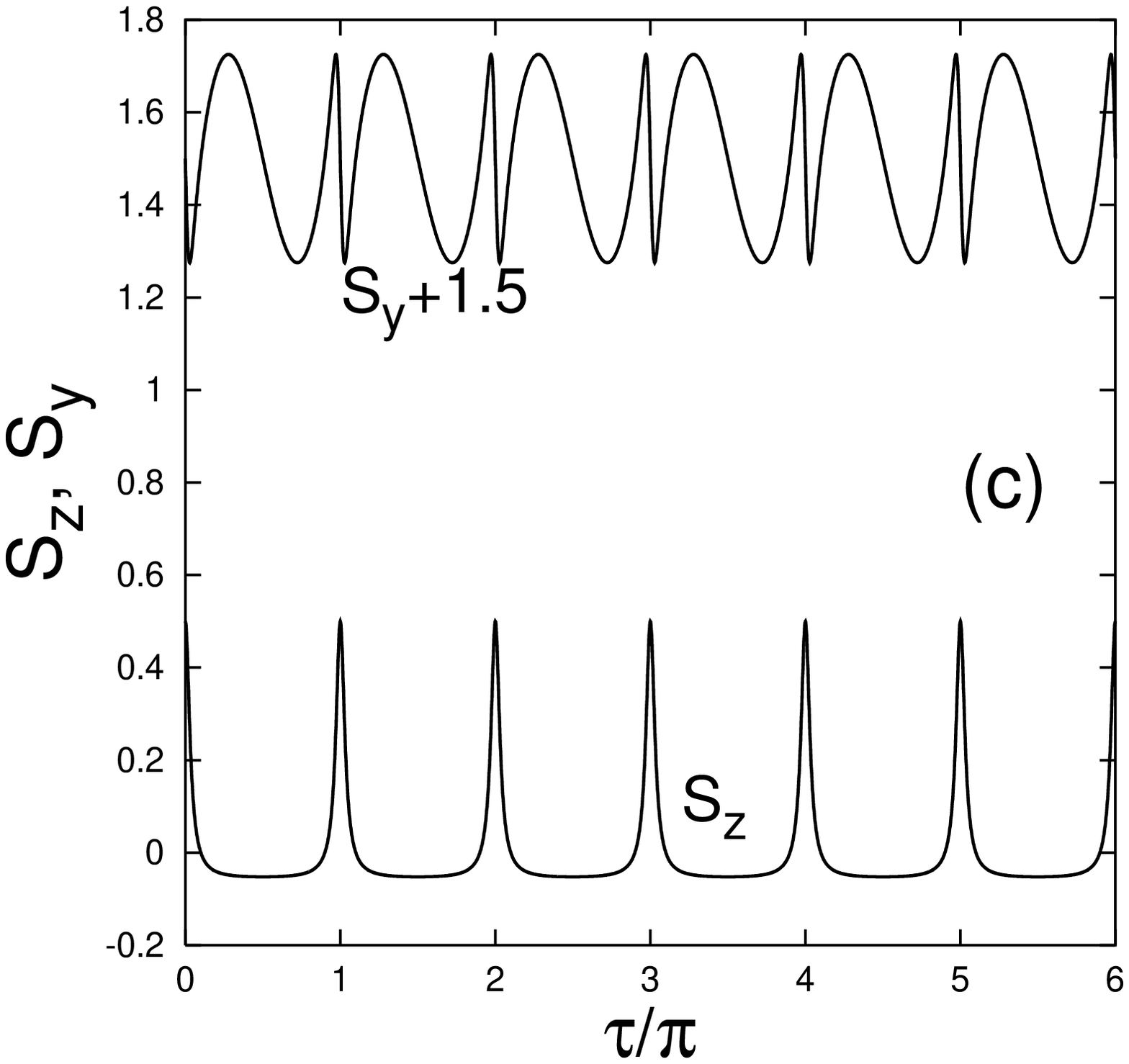,width=7.5cm,height=7.5cm}
\end{center}

\newpage

\begin{center}{\Large\bf Rao \& Xie: FIG.2/PHYSA}\end{center}
 
\vspace{0.5cm}
 
\begin{center}
 
\epsfig{file=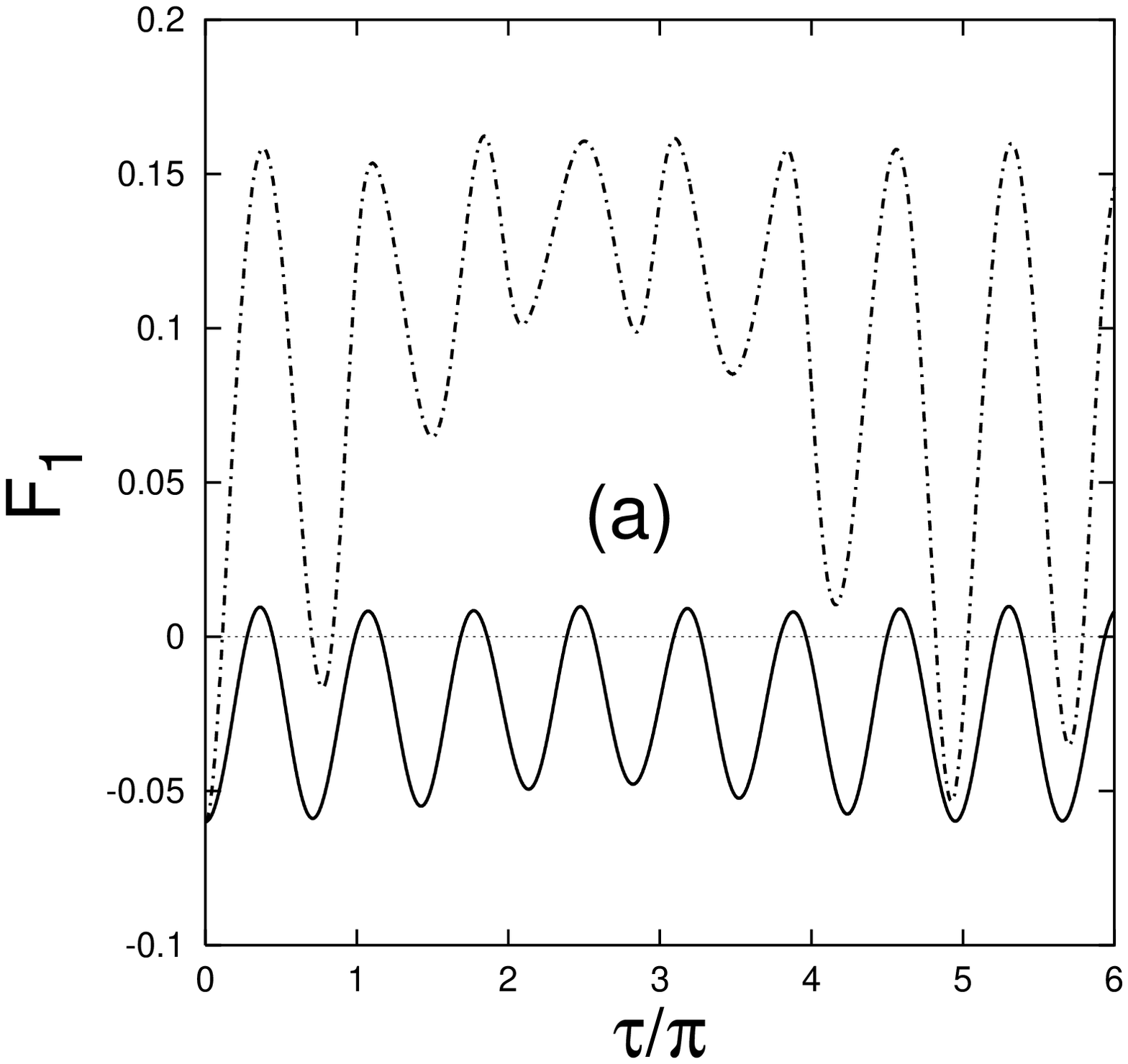,width=7.5cm,height=7.5cm}
\epsfig{file=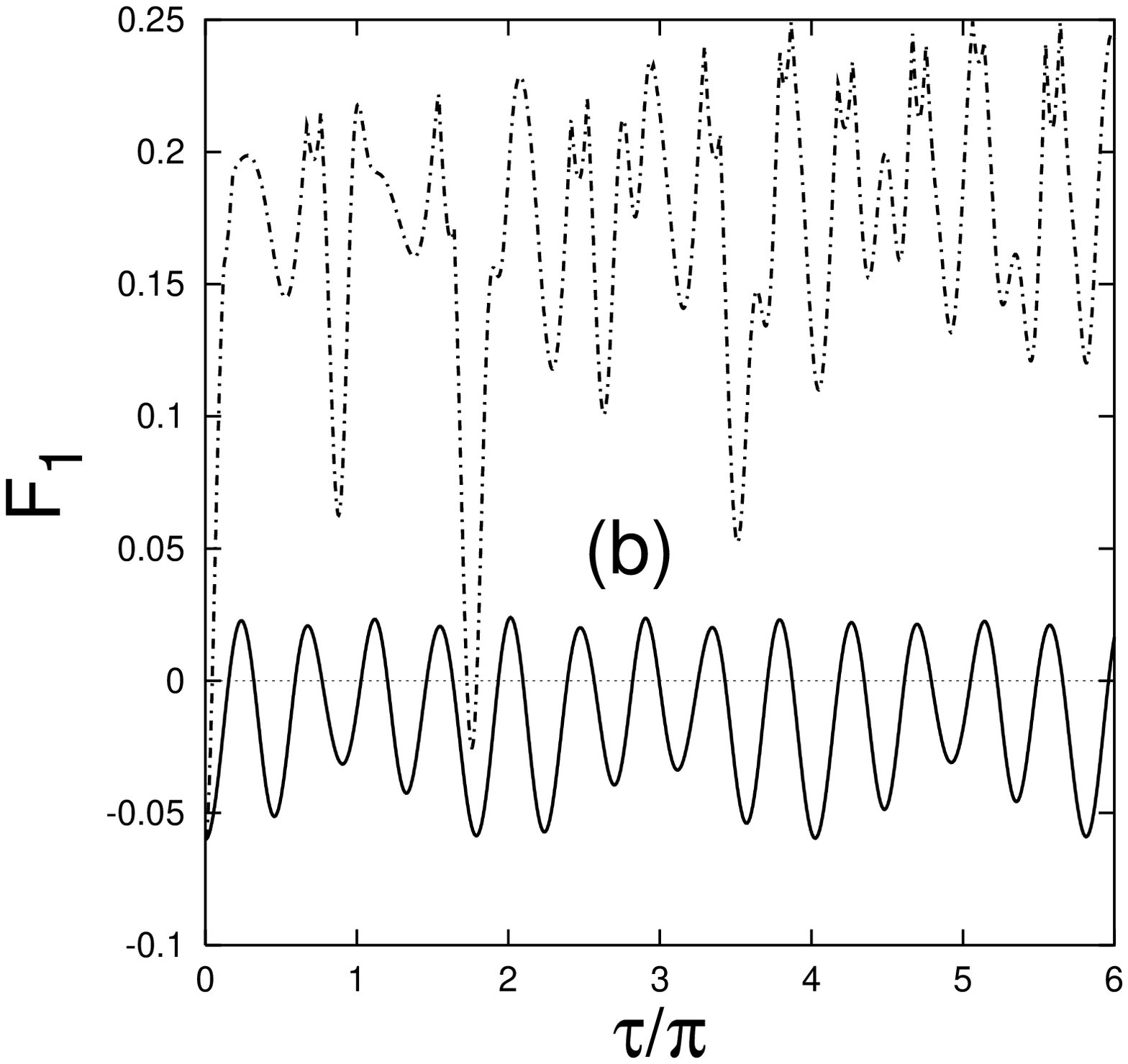,width=7.5cm,height=7.5cm}
 
\vspace{0.5cm}
 
\epsfig{file=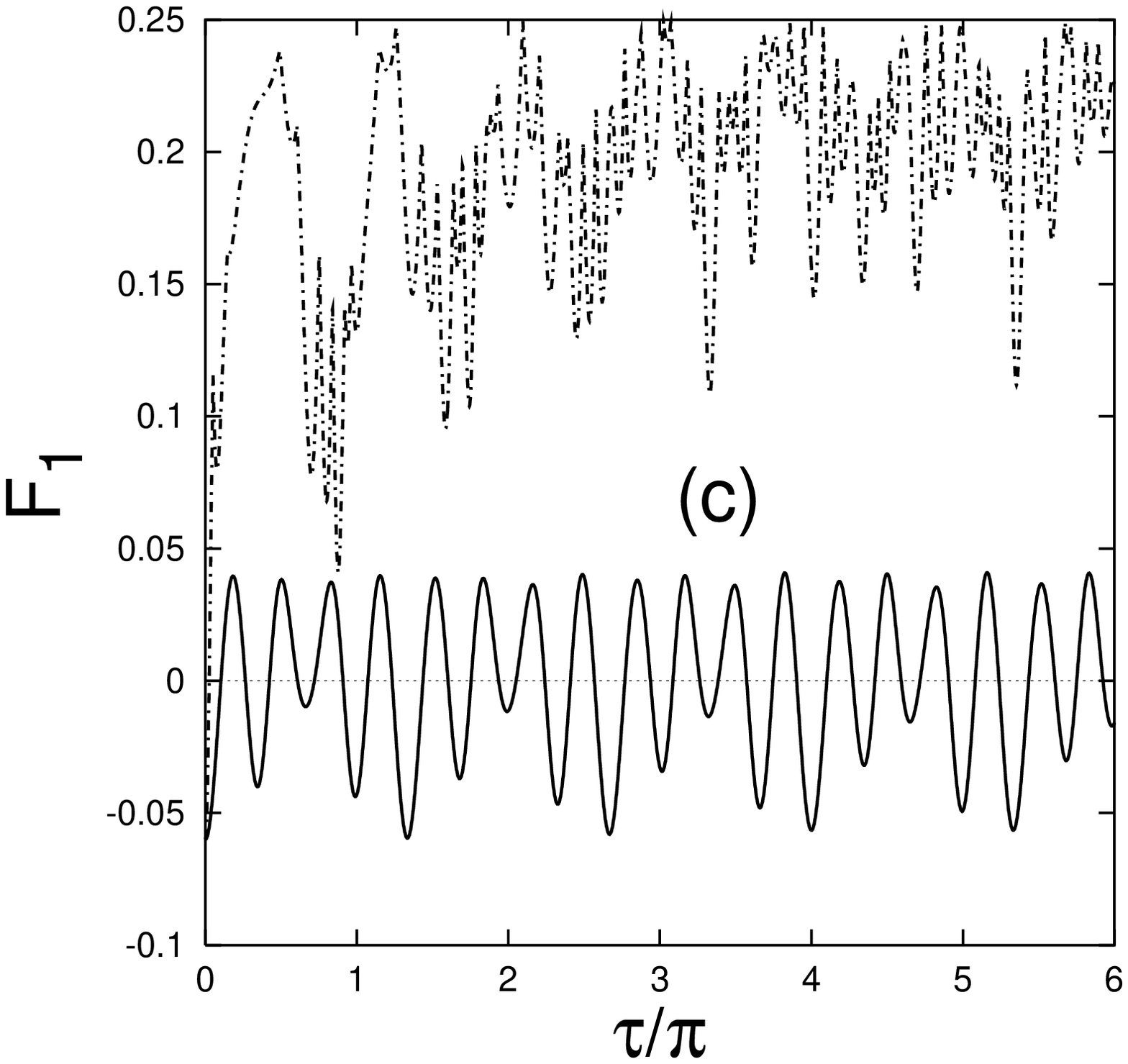,width=7.5cm,height=7.5cm}
\epsfig{file=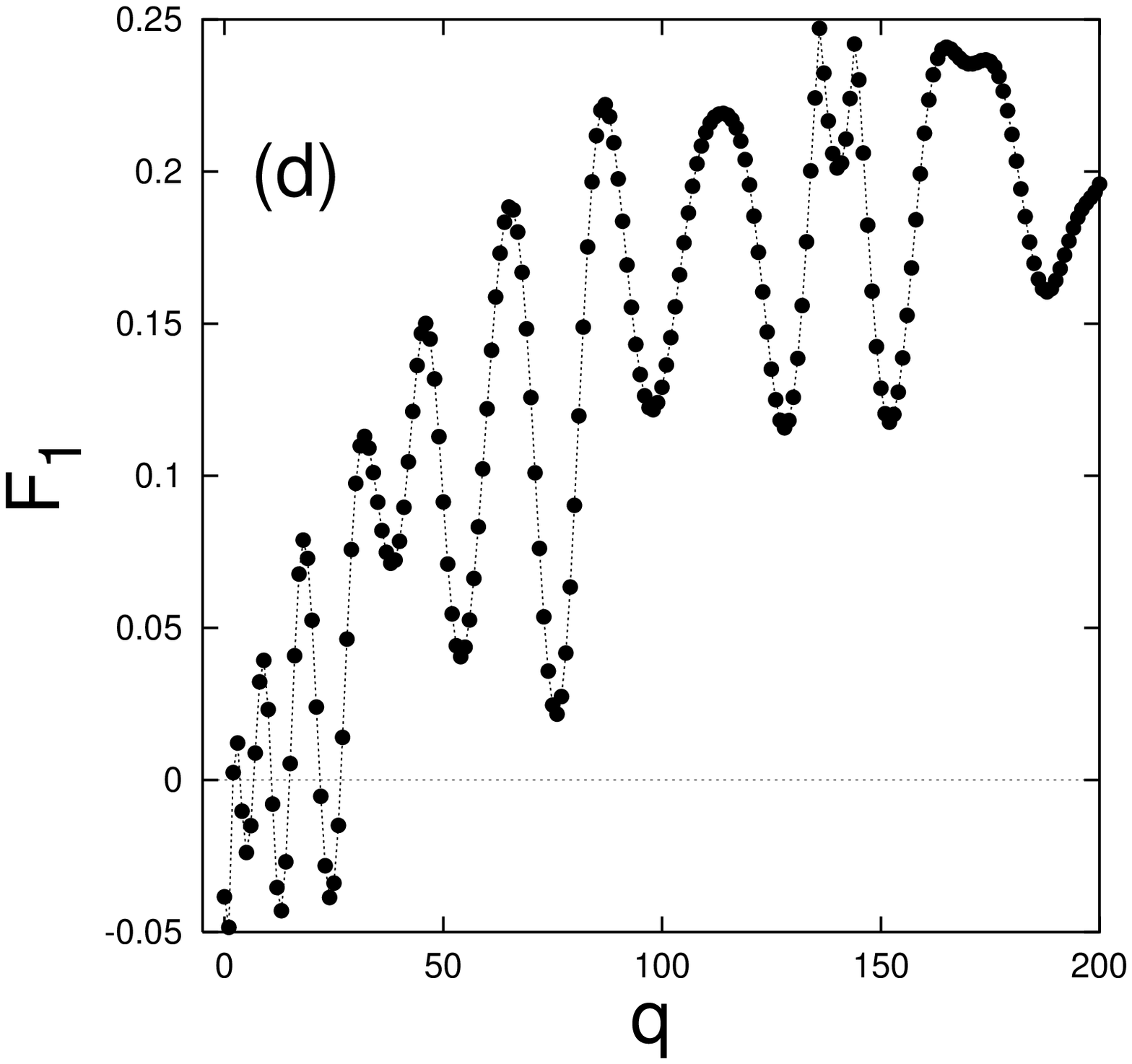,width=7.5cm,height=7.5cm} 
\end{center}

\newpage
 
\begin{center}{\Large\bf Rao \& Xie: FIG.3/PHYSA}\end{center}
 
\vspace{3cm}
 
\begin{center}
 
\epsfig{file=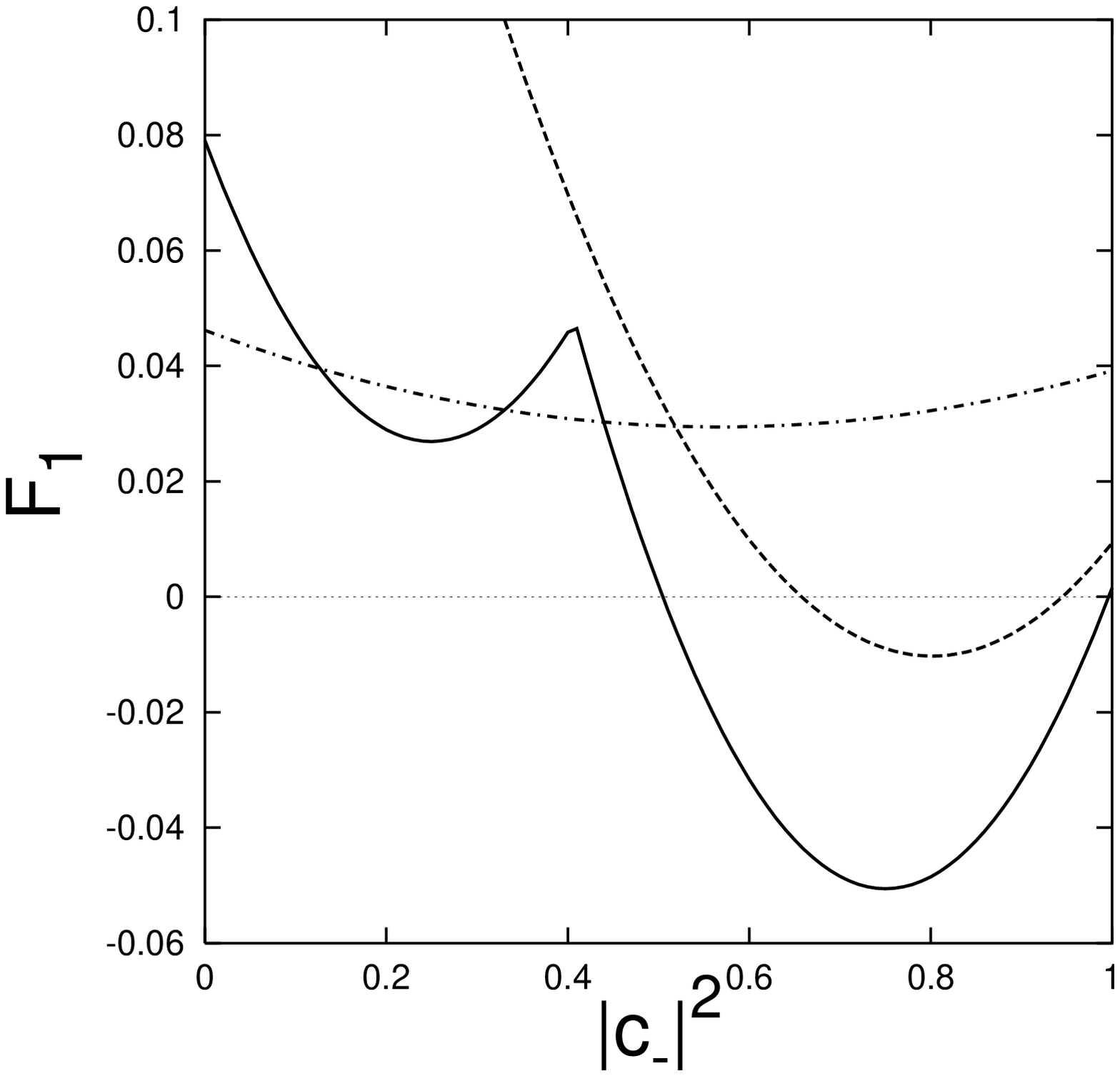,width=11cm,height=11cm}
 
\end{center}

\newpage
 
\begin{center}{\Large\bf  Rao \& Xie: FIG.4/PHYSA}\end{center}
 
\vspace{3cm}
 
\begin{center}
 
\epsfig{file=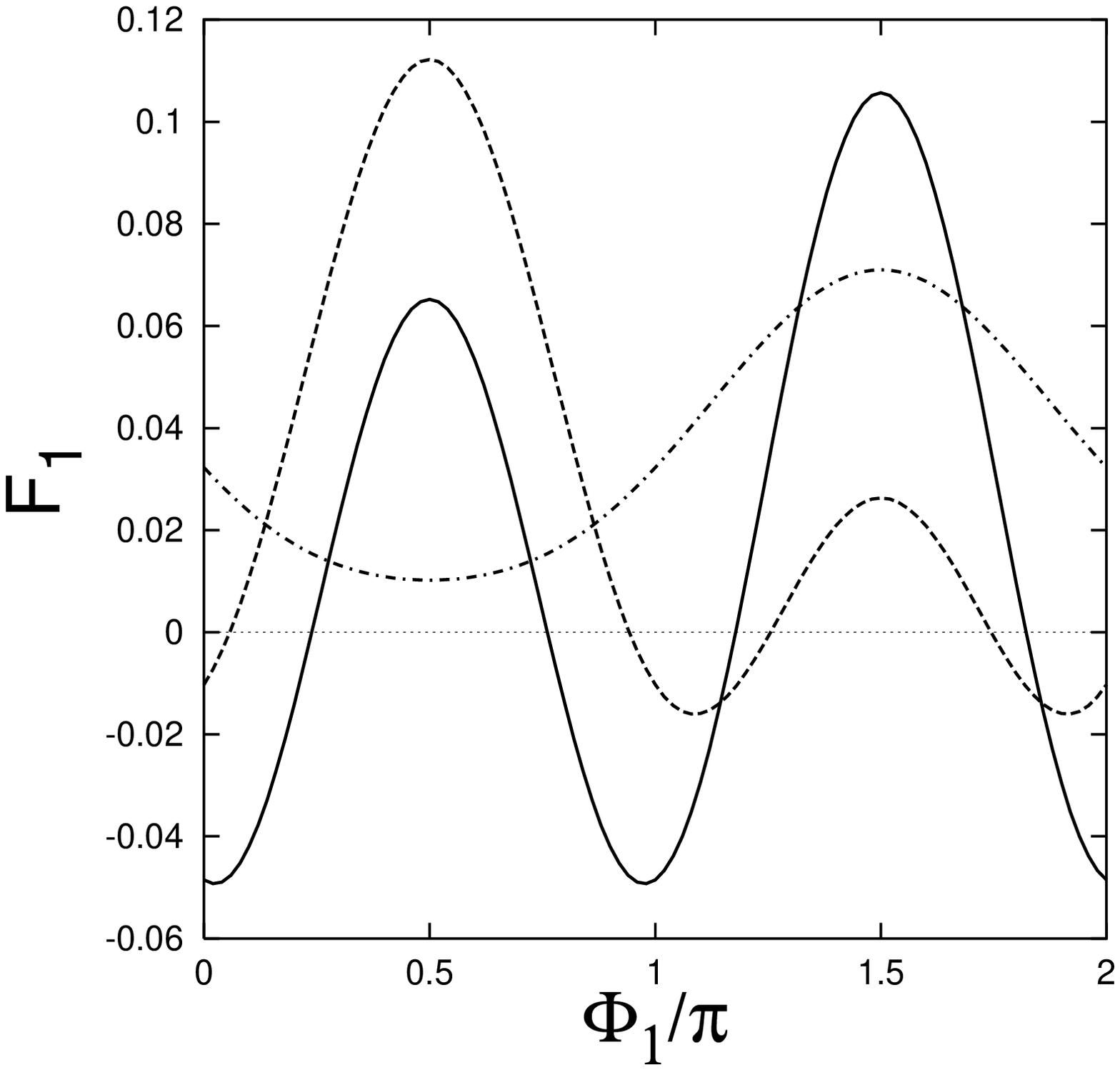,width=11cm,height=11cm}
 
\end{center}

\newpage

\begin{center}{\Large\bf Rao \& Xie: FIG.5/PHYSA}\end{center}
 
\vspace{0.5cm}
 
\begin{center}
 
\epsfig{file=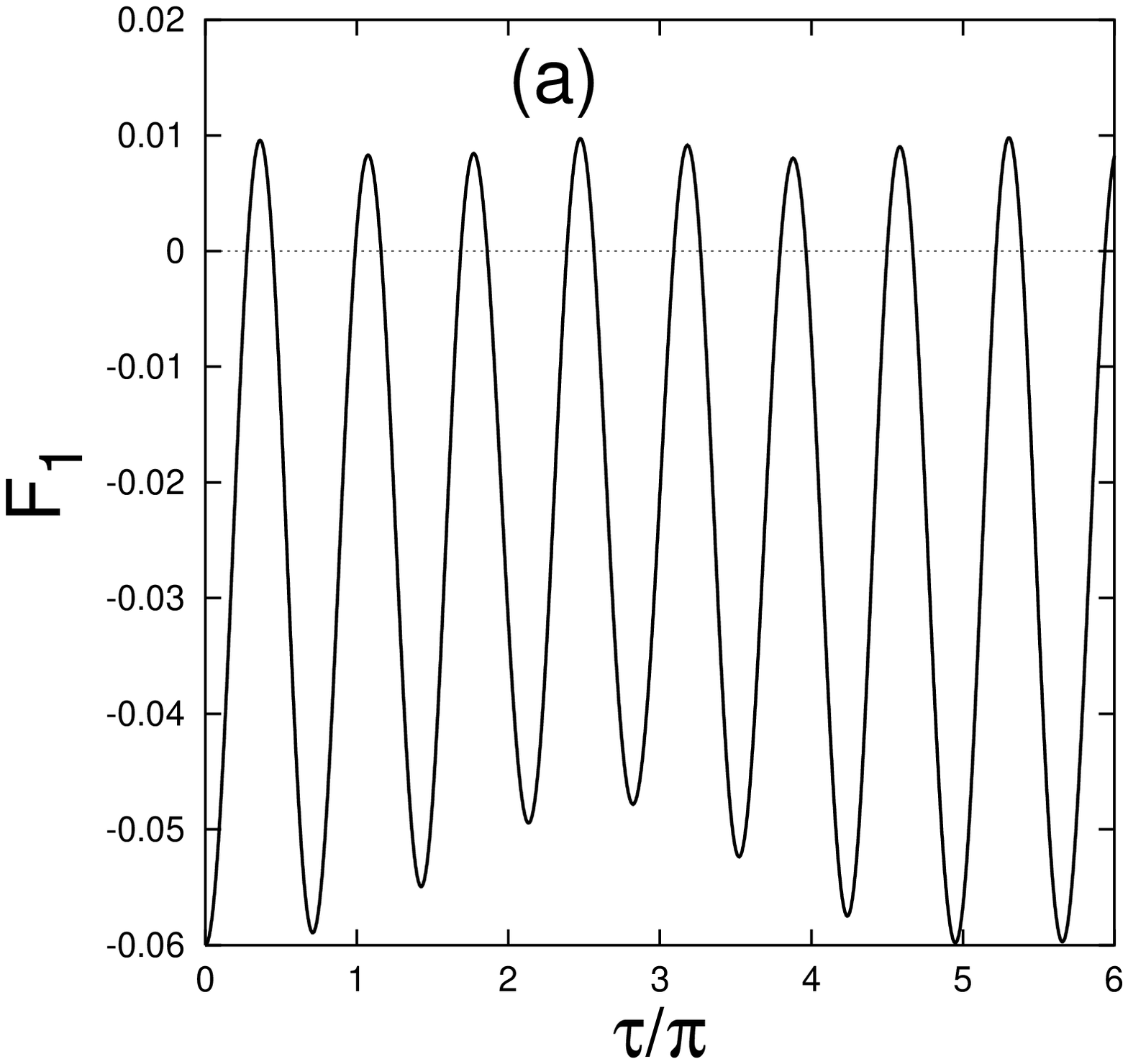,width=7.5cm,height=7.5cm}
\epsfig{file=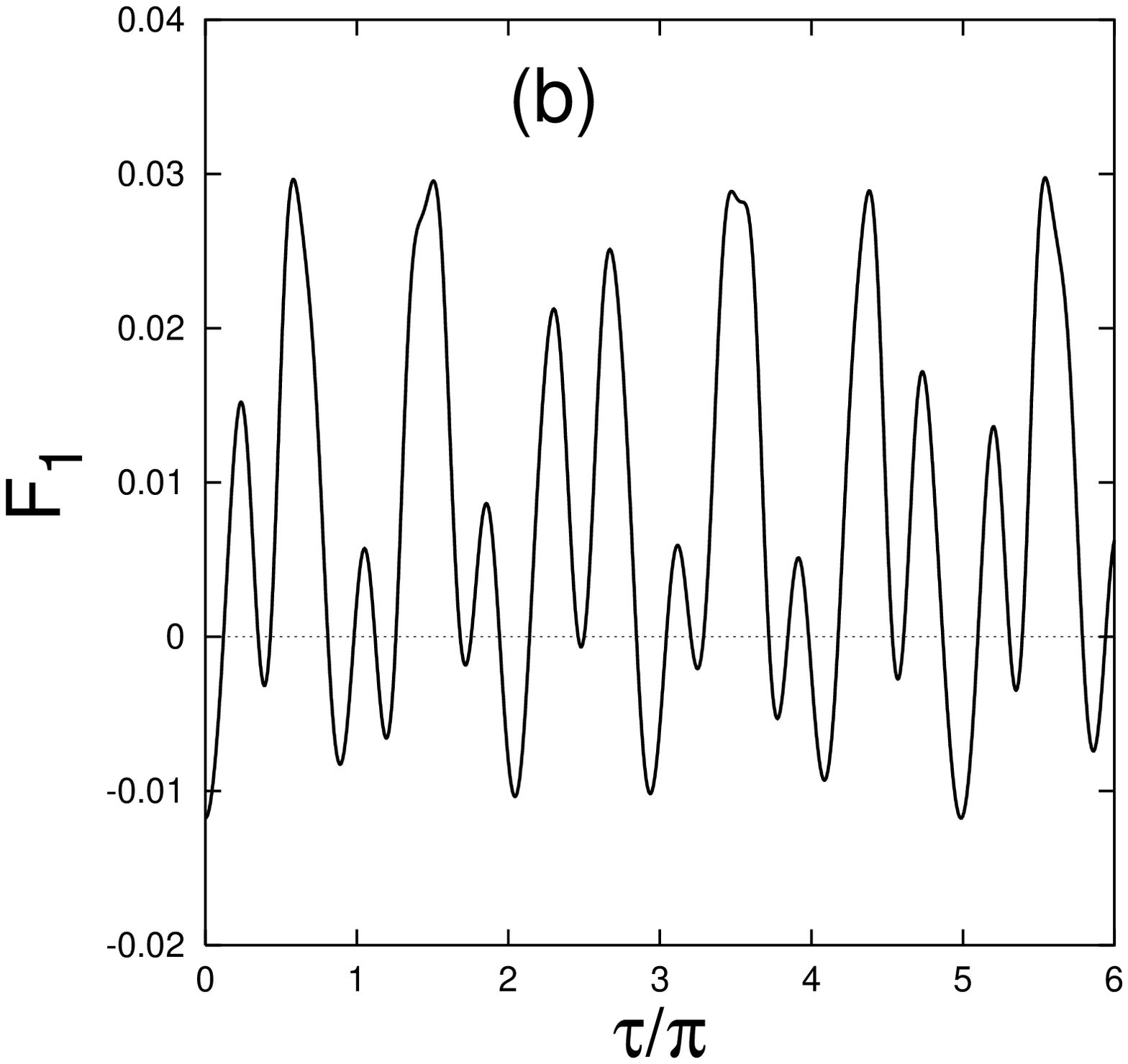,width=7.5cm,height=7.5cm}
 
\vspace{0.5cm}

\epsfig{file=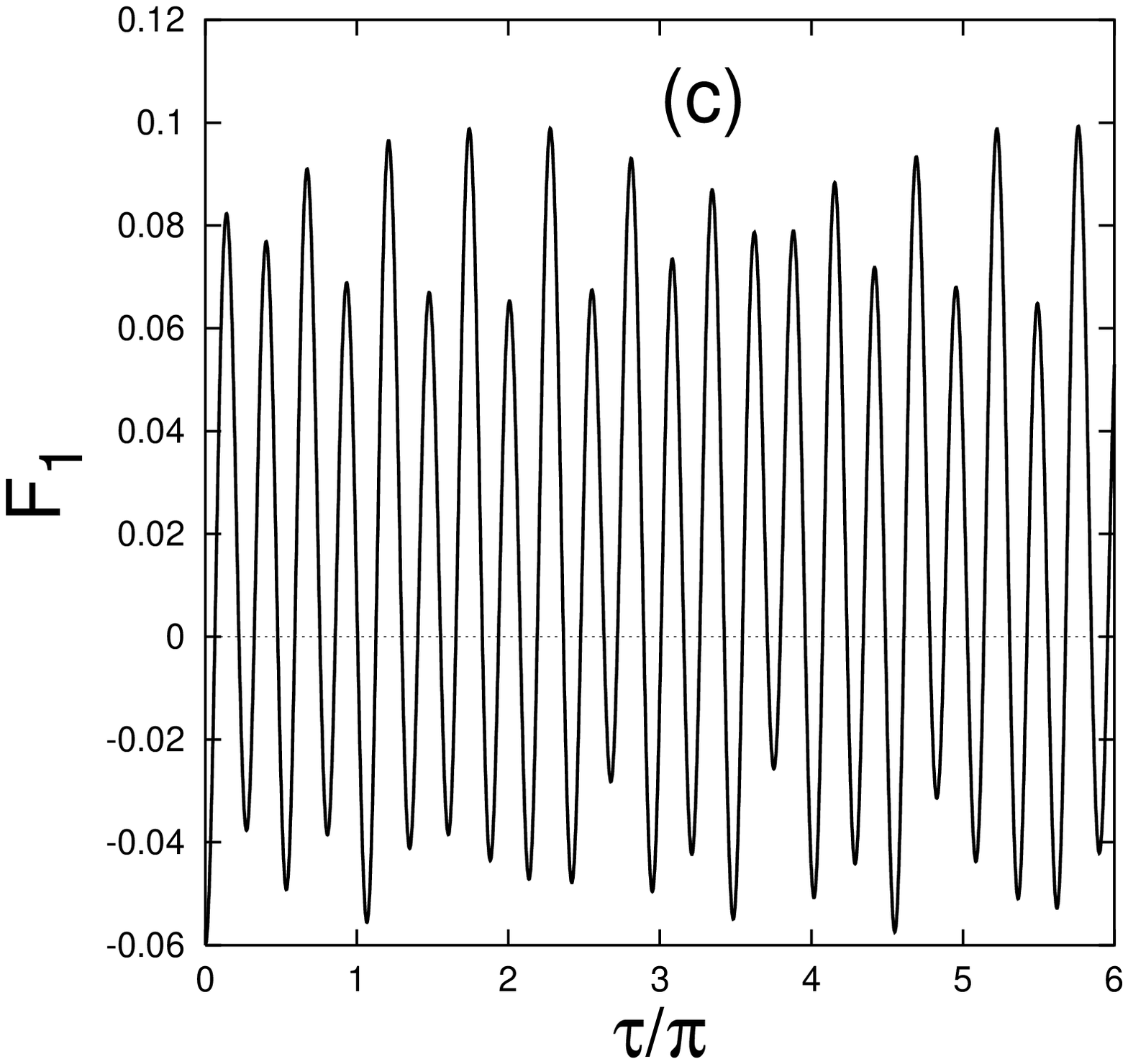,width=7.5cm,height=7.5cm}
\epsfig{file=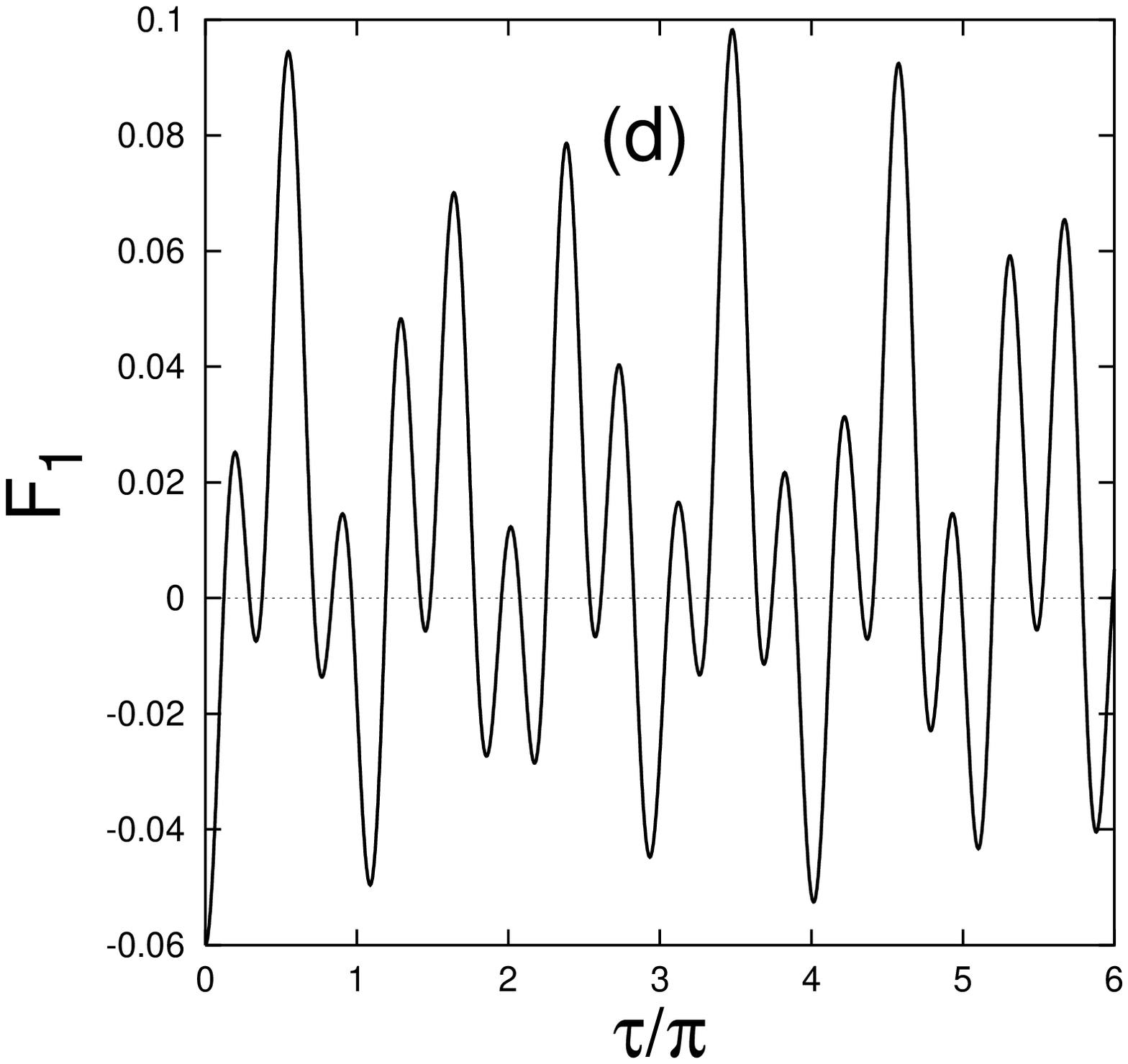,width=7.5cm,height=7.5cm}
 
\end{center}

\newpage
 
\begin{center}{\Large\bf Rao \& Xie: FIG.6/PHYSA}\end{center}
 
\vspace{3cm}
 
\begin{center}
 
\epsfig{file=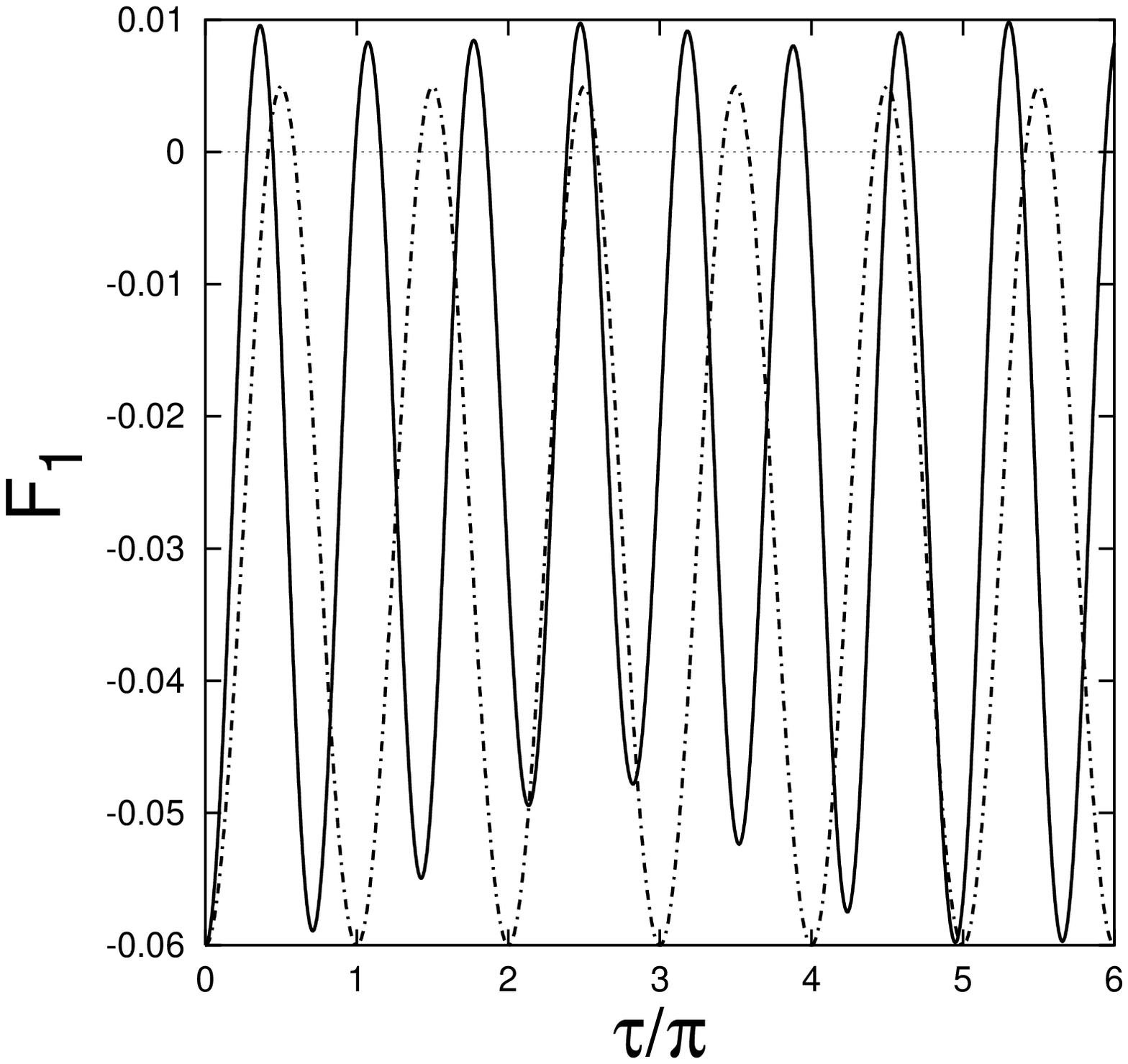,width=15cm,height=11cm}
 
\end{center}

\end{document}